\documentclass[twocolumn,article]{aastex701}
\usepackage{subcaption}
\usepackage{graphicx}	
\usepackage{amsmath,amssymb}
\usepackage{cleveref}
\usepackage[T1]{fontenc}
\usepackage{natbib}
\usepackage{float}
\usepackage{graphicx}
\usepackage{dcolumn}
\usepackage{amsmath}
\usepackage{amssymb}
\usepackage{multirow}
\usepackage{bm}
\usepackage{color}
\usepackage{xcolor}
\usepackage[normalem]{ulem}
\usepackage[dvipsnames]{xcolor}
\usepackage{hyperref}

\graphicspath{{./}{figures/}}

\begin{document}

\title{Low-angular-momentum accretion shocks can power weak-to-moderate X-ray flares from Sgr~A*}

\author[orcid=0000-0002-5730-0376]{Samik Mitra}
\affiliation{International Center for Theoretical Sciences, Tata Institute of Fundamental Research, Bangalore, India}
\affiliation{Center for Theoretical Physics, Polish Academy of Sciences, Al. Lotników 32/46, 02-668 Warsaw, Poland}
\affiliation{Department of Theoretical Physics and Astrophysics, Masaryk University, Kotl\'{a}\v{r}sk\'{a} 267/2, 611 37 Brno, Czech Republic}

\email[show]{samik.mitra@icts.res.in}

\author[orcid=0000-0001-5848-4333, gname='Bo\.zena',sname='Czerny']{Bo\.zena Czerny}
\affiliation{Center for Theoretical Physics, Polish Academy of Sciences, Al. Lotników 32/46, 02-668 Warsaw, Poland}
\email[show]{bcz@cft.edu.pl}

\author[orcid=0000-0001-6450-1187]{Michal Zaja\v{c}ek}
\affiliation{Department of Theoretical Physics and Astrophysics, Masaryk University, Kotl\'{a}\v{r}sk\'{a} 267/2, 611 37 Brno, Czech Republic}
\email[noshow]{zajacek@mail.muni.cz}

\author[orcid=0009-0004-8539-3516]{Zach Sumners}
\affiliation{McGill University, Montreal QC H3A 0G4, Canada}
\affiliation{Trottier Space Institute, 3550 Rue University, Montreal, Quebec, H3A 2A7, Canada}
\email[]{ronald.sumners@mail.mcgill.ca}

\begin{abstract}
X-ray flares from Sgr~A* span broad ranges in duration, fluence, and luminosity, but their origin remains unsettled. We test whether standing shocks in low-angular-momentum magnetized accretion flows can provide a viable energy reservoir to account for these events. Using semi-analytic trans-magnetosonic shock solutions, we estimate the kinetic energy available in the downstream post-shock flow and compare it with the 25-year \textit{Chandra} X-ray flare catalog. For each theoretical solution, we compute the required efficiency $\epsilon=E_{\rm data}/E_{\rm sh}$, where $E_{\rm data}$ is the observed radiated energy for each flare and $E_{\rm sh}$ is the available energy in the shocked flow. Notably, the weak flares require $\epsilon\sim10^{-3}$--$10^{-2}$, while moderate flares require a few percent. We perform the analyses for weakly and highly spinning cases, and the resulting weak-to-moderate flare energy budget remains unchanged. For the fiducial accretion rate and bolometric correction, the kinetic energy reservoir in standing shock is sufficient for the observed weak-to-moderate flare population, whereas the strongest events likely require higher efficiency which can be mediated via additional magnetic energy dissipation channel.
\end{abstract}


\section{Introduction}

The Galactic center hosts a variable compact radio source, Sgr~A*, which is associated with a supermassive black hole (SMBH) of mass $\sim 4\times10^{6}\,M_{\odot}$ \citep{gravity2022}. Despite being the nearest SMBH ($\sim8\,\mathrm{kpc}$; \citealt{2010RvMP...82.3121G,2017FoPh...47..553E,2022RvMP...94b0501G}), it is extraordinarily underluminous: its bolometric luminosity is only $\sim10^{-9}$ of the Eddington value \citep{1998ApJ...492..554N}. Sgr~A* is detected across the electromagnetic spectrum with its spectral energy distribution (SED) peaking in the millimeter/submillimeter domain \citep{1974ApJ...194..265B,2013CQGra..30x4003F,2021ApJ...917...73W}. The most direct information about the horizon-scale flow structure comes from the \cite{2022ApJ...930L..12E,2024ApJ...964L..25E} images, while time-domain observations reveal persistent flaring activity across all wavebands \citep{2012A&A...537A..52E,Yuan-etal2018,2021ApJ...917...73W} whose physical origin remains debated. Previous variability studies have characterized Sgr~A* across radio \citep{Bower2015RadioMillimeter,Chen2023LLAGNVar}, submillimeter \citep{Subroweit2017SubmmRadio,Michail2021Multiwavelength}, infrared \citep{Genzel2003NIRFlares,Ghez2004VariableIR,Hora2014SpitzerIRAC,Do2019BrightNIR,2025ApJ...979L..20V,Michail2026MIR}, and X-ray wavelengths \citep{Baganoff2001RapidXray,Nowak2012BrightestFlare,Neilsen2013Census,Haggard2019BrightestFlares,Bouffard2019NoSignG2}.

There are, however, several clear observational properties that every complete flare model should meet, namely that for simultaneous multi-wavelength observations, every X-ray flare has an associated infrared counterpart but not vice versa \citep{2006ApJ...644..198Y,2008A&A...479..625E,2009ApJ...698..676D,dodds_eden2010,2011ApJ...728...37D,2017MNRAS.468.2447P,2021A&A...654A..22G,2025ApJ...980L..35Y,2025ApJ...979L..20V}. In contrast to the infrared emission, which exhibits a well-characterized variable low-flux component, the X-ray emission is dominated by discrete flares superimposed on the faint quiescent component \citep{Baganoff2001RapidXray, Neilsen2013Census,2019ApJ...871..161B}. In addition, the bright multi-wavelength flares are often described by a broken power-law SED between the infrared and X-ray domains, consistent with synchrotron emission with a cooling break \citep{2017MNRAS.468.2447P,2021A&A...654A..22G,Michail2026MIR}, with some flares having time-delayed counterparts towards the submm/mm/radio domains \citep{2008A&A...492..337E,2008ApJ...682..361Y,2022ApJ...930L..19W,Michail2026MIR}. Finally, the infrared flux distribution is of a power-law nature for high fluxes (flares) while it has a log-normal distribution for low fluxes with the median value of $\sim 1.1\,{\rm mJy}$ \citep{2020A&A...638A...2G}.  

Current interpretations of the Sgr~A* variability are largely studied with the three-dimensional (3D) general relativistic MHD (GRMHD) simulations of accretion flows, particularly in the magnetically arrested disk \citep[MAD;][]{Narayan-etal2003} and standard and normal evolution \citep[SANE;][]{Narayan-etal2012} paradigms. In these models, flares are commonly associated with magnetic reconnection events, turbulent plasma fluctuations, or orbiting hot spots \citep{dodds_eden2010,Scepi-etal2022,Nathanail-etal2022,Ripperda-etal2022}. While such scenarios naturally produce variability, establishing simple connections between flare observables and the underlying flow properties remains challenging.

An alternative possibility is provided by low-angular-momentum accretion flows \citep{abramowicz1981,Fukue1987,Chakrabarti1989,Chakrabarti1990,Takahashi-etal2002,Takahashi-etal2006,Das-etal2009,Das2012,2015MNRAS.447.1565S,Sukova-etal2017,Dihingia-etal2019,Nazari2024,mitra2024}. Here, the competition between gravity and centrifugal support can create a “virtual” centrifugal barrier that slows the convergent matter. When the supersonic flow undergoes a standing shock, it is abruptly decelerated and compressed, and part of its radial bulk kinetic energy is transferred to the thermal energy of the downstream plasma. The resulting post-shock flow is therefore hotter, denser, and geometrically thicker than the corresponding smooth solution, forming the post-shock corona (PSC). The PSC is thus the hot, advecting downstream flow produced by the shock and provides a potential site for particle energization and hard-X-ray production \citep{Chakrabarti-Titarchuk1995,Le-Becker2005,Becker-etal2008,Becker-etal2011}. Recent GRMHD simulations have demonstrated that such shocks can persist in low angular momentum magnetized flows, although their survival depends on the field configurations and selective ranges of the flow parameters \citep{mao2025,Dihingia-etal2025,Dihingia-etal2026}.

Whether Sgr~A* operates in this regime depends on the angular momentum of the accreting gas supplied by the winds of young, massive OB stars orbiting at $\sim0.04$--$0.5\,\mathrm{pc}$ from Sgr~A* \citep{2004ApJ...613..322Q,2008MNRAS.383..458C,2010ApJ...716..504S,2015MNRAS.453..775G,2018MNRAS.479.4778Y}. Although numerical simulations suggest substantial angular momentum at large radii \citep{ressler2018,Ressler-etal2020}, they also predict strong outflows that preferentially remove high-angular-momentum material. The gas that ultimately reaches the BH may therefore possess significantly lower angular momentum than the average stellar-wind value. We note that the signature of a hot wind from the Sgr~A* circumnuclear environment was identified based on the elongated structure characterized by the decrease in the molecular-gas content \citep[$\sim 1\,{\rm pc}$ long and with the $\sim 45^{\circ}$ opening angle, see][and references therein]{2026ApJ..1004L...7G}, which is filled with hot X-ray emitting gas. Another signatures of the nuclear outflow include stellar bow-shock sources (X3, X7, X8) pointed away from Sgr~A* \citep{2010A&A...521A..13M,2016MNRAS.455.1257Z,2019A&A...624A..97P,2020MNRAS.499.3909Y}, a significantly blue-shifted ionized gas component within $\sim 0.2$ pc \citep{2019ApJ...872....2R,2020MNRAS.499.3909Y}, and a very broad ($\pm 500\,{\rm km\,s^{-1}}$) red-/blue-shifted double-peak recombination-line spectral feature on milliparsec scales from Sgr~A* \citep{2019Natur.570...83M,2020MNRAS.499.3909Y}.

In this manuscript, we investigate whether the shocked low-angular-momentum magnetized accretion flow contains a sufficiently large energy reservoir to account for weak-to-moderate X-ray flares from Sgr~A*. Using the semi-analytic relativistic trans-magnetosonic models of \citet{mitra2022}, \citet{mitra2024}, and \citet{Mitra2025}, we construct an order-of-magnitude estimate of the kinetic-energy reservoir available at the shock front and compare it with the recent 25-year \textit{Chandra} flare census of \citet{Chandra2026} (see also \citet{Neilsen2013Census,2015ApJ...799..199N,2015MNRAS.454.1525P,Yuan-etal2018,Bouffard2019NoSignG2} for previous Sgr~A* X-ray flare population and timing studies). We additionally compare the infall and sound-wave crossing timescales and do the analysis for a weakly spinning BH, $a_{\rm k}=0.5$ and a highly rotating BH with $a_{\rm k}=0.94$. Rather than attempting to reproduce individual flare light curves or spectra, we perform an energy-budget test by computing the radiative efficiency required for each observed flare.

\section{Model}\label{sec:model}

We model the accretion onto Sgr~A* ($M_{\rm BH}=4.2\times10^{6}\,M_{\odot}$) as a stationary flow, motivated by the approximately steady quiescent emission when averaged over timescales much longer than individual flares ($\gg 1$ hour). The spin of Sgr~A* is not yet determined uniquely. We therefore treat $a_{\rm k}=0.94$ as a fiducial choice rather than as an observationally established spin measurement \citep{Daly2019,EHT2023}. In general, a moderate to high prograde spin value is preferred based on the comparison of horizon-scale GRMHD models as well as Sgr~A* flare calculations with observations \citep{Genzel2003NIRFlares,2011MNRAS.413..322Z,2018ApJ...863...15W,2018acps.confE..48E,2022ApJ...930L..16E,2024ApJ...964L..26E} although specific spin values are highly model-dependent. In contrast, the anisotropic dynamical configuration of S stars \citep{2020ApJ...896..100A,2020ApJ...899...50P} seems to prefer a low spin \citep{2020ApJ...901L..32F}. Future independent constraints on the Sgr~A* spin will be provided by monitoring S stars with the smallest pericenter distances \citep{2020ApJ...899...50P,ElDayem2026} based on the Lense-Thirring precession of their line of nodes. We therefore repeat the principal calculations for a lower-spin case, $a_{\rm k}=0.5$, to assess the sensitivity of the flare comparison to spin. The stationary, axisymmetric ideal-GRMHD equations are solved in the exact Kerr spacetime in Boyer--Lindquist coordinates. In this work, we express the length $r$ and time $t$ in terms of $r_{\rm g}$ and $r_{\rm g}/c$, $r_{\rm g} = G M_{\rm BH}/c^2$ being the gravitational radius, where $G$ is the gravitational constant and $c$ is the speed of light. The quasi-steady flow contains a low-angular-momentum, axisymmetric accretion disk threaded by both radial ($b^r$) and toroidal ($b^\phi$) magnetic field components. Furthermore, the convergent flow is assumed to be confined near the mid-plane, and therefore we choose $u^\theta\sim 0$ and we additionally set $b^\theta\sim 0$. With these assumptions, we solve the GRMHD conservation equations,
\begin{equation}
    \nabla_\mu(\rho u^\mu) = 0, \qquad
    \nabla_\mu T^{\mu\nu} = 0, \qquad
    \nabla_\mu {^*\!F}^{\mu\nu} = 0,
    \label{eq:grmhd}
\end{equation}
where $\rho$ is the rest-mass density, $u^\mu$ the four-velocity, $T^{\mu\nu}$ the total (fluid + magnetic) stress-energy tensor, and ${^*\!F}^{\mu\nu}$ is the dual Faraday tensor. Under ideal-MHD conditions ($u_\mu b^\mu=0$, where $b^\mu$ is the magnetic field in the fluid-frame) the magnetic field is frozen into the plasma, and the flow is fully characterized by four globally conserved quantities: energy flux $\mathcal{E}(=-T^r_t/\rho u^r)$, angular momentum flux $\mathcal{L}(=T^r_\phi/\rho u^r)$, radial magnetic flux $\Phi(=r^2 \{u^r b^t-u^t b^r\})$, and the iso-rotation parameter $F(=r^2 \{u^r b^\phi-u^\phi b^r\})$. We adopt the relativistic equation of state \citep{Chattopadhyay-Ryu2009}, which self-consistently accounts for the radial dependence of the adiabatic index, $\Gamma(r)$. 

Following \cite{mitra2024}, we first perform the magnetosonic point analysis, and eventually solve the wind equations to calculate various flow parameters, $u^r,u^\phi,b^r,b^\phi,\rho$, and pressure $p$. For a given parameter set $(\mathcal{E}, \mathcal{L}, \Phi, F, a_{\rm k})$, the flow either passes smoothly through a single sonic point (shock-free solution; see Figure~\ref{fig:two_examples}a) or possesses two saddle-type sonic points -- an outer $r_\mathrm{out}$ and an inner $r_\mathrm{in}$ -- in which case a centrifugal pressure-supported standing shock forms at $r_\mathrm{sh}$ (see Figure~\ref{fig:two_examples}b). The shock location is determined by the following jump conditions \citep{Takahashi-etal2002},

\begin{equation}
\begin{split}
    \bigl[\rho u^r\bigr] = 0, \quad
    \bigl[T^r{}_t\bigr] = 0, \quad
    \bigl[T^r{}_\phi\bigr] = 0, \\
    \bigl[T^r{}_r\bigr] = 0, \quad
    \bigl[{^*\!F}^{rt}\bigr] = 0, \quad
    \bigl[{^*\!F}^{r\phi}\bigr] = 0,
\end{split}
    \label{eq:RH}
\end{equation}
where $[\cdots]$ denotes the difference between post- and pre-shock quantities.  Note that the low-angular-momentum ($\lambda=-u_\phi/u_t$\footnote{$\lambda$ is the flow angular momentum.}) flow demands $\lambda\ll \lambda_{\rm Kep}$, where $\lambda_{\rm Kep}$ is the Keplerian angular momentum.

\section{Results}

\begin{figure}[h!]
\includegraphics[width=\columnwidth]{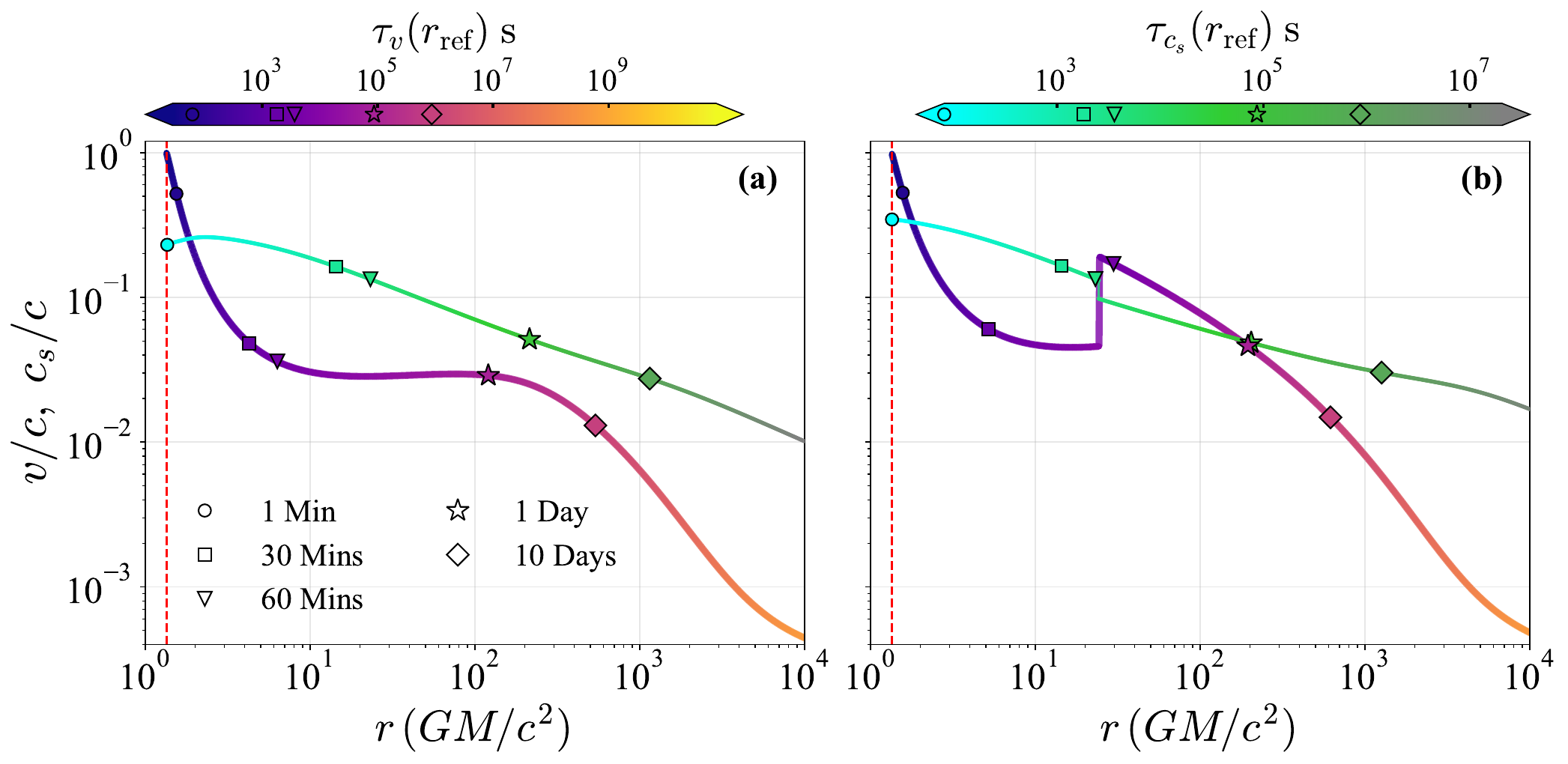}
\caption{Profiles of the dimensionless radial velocity $|v|$ (thick curve) and sound speed $c_s$ (thin curve) for (a) the smooth shock-free solution and (b) a shocked solution. The radial-velocity and sound-speed curves are color-coded by the cumulative infall/advection timescale $\tau_v(r_{\rm ref})$ and sound-wave crossing timescale $\tau_{c_s}(r_{\rm ref})$, respectively, in seconds. The vertical dashed line marks the horizon, and the symbols indicate reference radii for the characteristic times.}
\label{fig:two_examples}
\end{figure}

We first study global trans-magnetosonic, low-$\lambda$ flows around a Kerr BH accreting from the outer edge of the disk at $r_{\rm edge}=10^4$. The inflowing matter has a constant mass-accretion rate, $\dot{M}=10^{-8}M_{\odot}\,{\rm yr}^{-1}$. This value is used as a fiducial physical normalization, broadly consistent with near-horizon accretion-rate estimates for the quiescent flow of Sgr~A* \citep[see][and references therein]{2007ApJ...654L..57M,2010ApJ...716..504S}. For $\mathcal{L}=2.40$, the subsonic ($v<c_s$) convergent flow passes through a single magnetosonic point $r_{\rm in}=1.61$ (see Fig. \ref{fig:two_examples}a) and accretes smoothly onto the BH with supersonic speed ($v>c_s$) inside the magnetosonic point. We define the relativistic sound speed as $c_s^2=\Gamma p_{\rm gas}/(\rho h)$, where $p_{\rm gas}$ is the gas pressure and $h$ is the hydrodynamic specific enthalpy \citep[see][for details]{mitra2024}. For $\mathcal{L}=2.20$, the solution instead possesses multiple sonic points, at $r_{\rm in}=1.84$ and $r_{\rm out}=208.41$, and a corresponding standing shock at $r_{\rm sh}=24.19$ (see the vertical jump in Fig. \ref{fig:two_examples}b). We keep the other parameters\footnote{We define, $\Phi=\Phi_{13}\times10^{-13}$ and $F=F_{15}\times 10^{-15}$ in code units.} fixed at $\mathcal{E}=1.001$, $\Phi_{13}=10.50$, and $F_{15}=5$.
During the shock transition, the supersonic pre-shock (upstream) branch loses part of its bulk kinetic energy and jumps to a subsonic (downstream) branch before falling onto the BH. This pre-shock kinetic energy heats the post-shock branch, as also indicated by the jump in $c_s$ across the shock front in Fig.~\ref{fig:two_examples}b, and the shock compression eventually puffs up the inner disk. On the contrary, the smooth shock-free solution remains geometrically thinner and relatively cooler in this region. Hence, the post-shock region becomes the perfect site for the hot electrons to reside. Any soft photons that interact with these hot electrons can produce hard X-rays. This motivates us to test whether the X-ray flares of Sgr~A* can be powered from the available energy budget in the shocked flow.

However, as the model is stationary, we cannot predict the actual evolutionary timescales for this energy release. Still, we can introduce a local infall/advection timescale, $\tau_v(r_g/c) = \int_{r_{\rm h}+\delta r}^{r_{\rm ref}} dr/|v|$, where $|v|$ is the dimensionless inflow velocity, $r_{\rm h}$ is the horizon, and $\delta r = 0.001$. Here, $r_{\rm ref}$ is the reference radius from which the infall time for the flow to the horizon is evaluated. We notice from Fig.~\ref{fig:two_examples} that, over the range $r_{\rm ref}\simeq3$--$85$, this timescale spans $\tau_v \simeq4\times10^2$--$2\times10^4\,{\rm s}$, comparable to the observed duration range of Sgr~A* X-ray flares \citep{Chandra2026}. We also calculate the sound-wave crossing timescale, $\tau_{c_s}(r_g/c)=\int_{r_{\rm h}+\delta r}^{r_{\rm ref}}dr/c_s$, which overlaps the observed flare-duration range for $r_{\rm ref}\simeq5$--$70$, consistent with the characteristic acoustic timescale discussed by \citet{Okuda-etal2023}. We therefore use these overlaps only as local characteristic dynamical timescales for comparison with the flare durations. We use $\tau_{c_s}$ for the primary comparison below and retain $\tau_v$ as a complementary test in the Appendix; neither timescale is interpreted as an exact prediction of the observed flare duration. Given the timescale, we will eventually discuss how the corresponding shock-induced flow will be able to cater to the energy budget (fluence) for individual flares.

\begin{figure}[h!]
\centering
\includegraphics[width=\columnwidth]{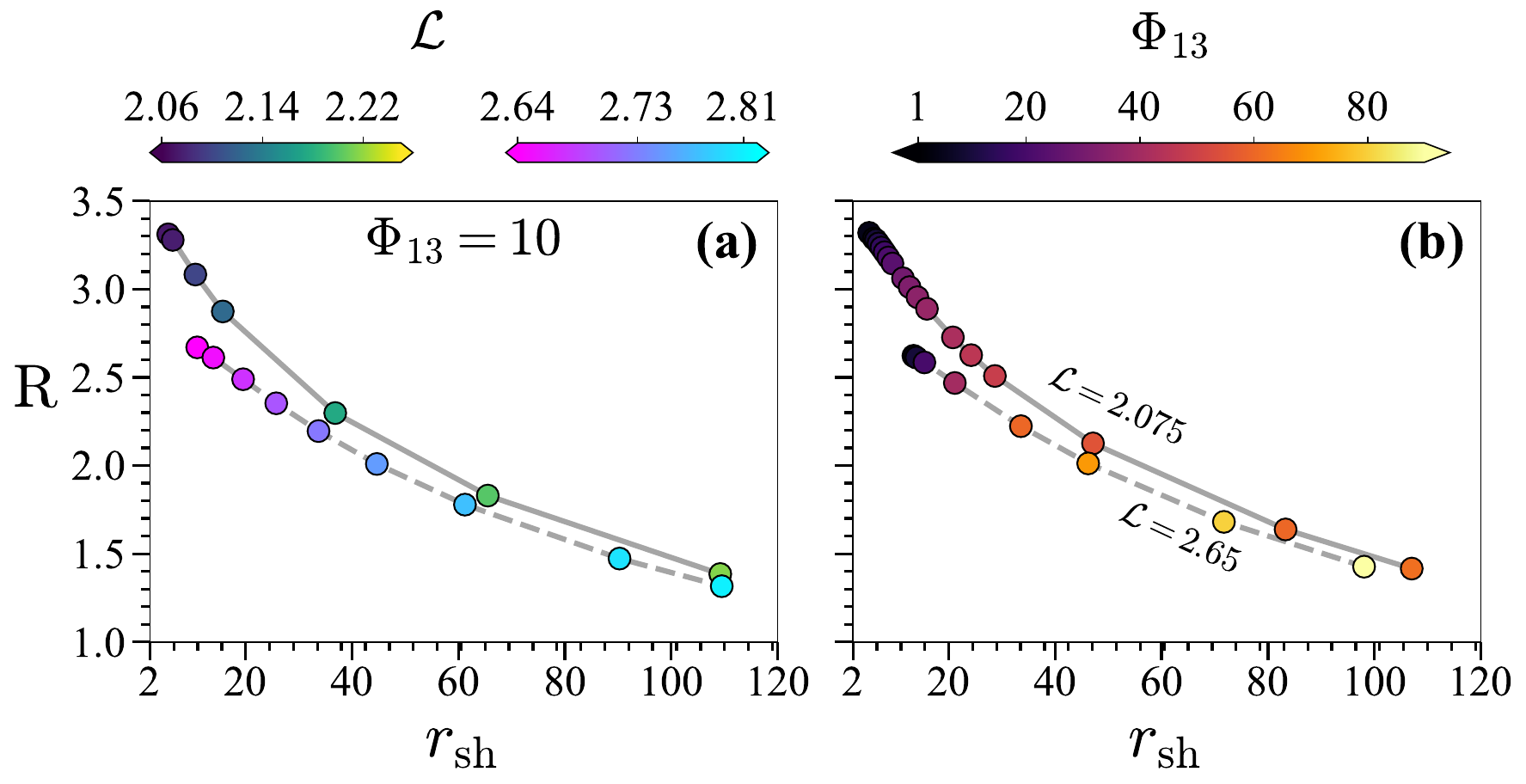}
\caption{We plot the shock location ($r_{\rm sh}$) and the corresponding compression ratio $R$ for the shocked solutions with varying (a) $\mathcal{L}$, (b) $\Phi_{13}$, keeping the other parameters fixed. In both panels we fix $\mathcal{E}=1.001$ and $F_{15}=5$. The colorbar in each panel represents the range of variations of the respective quantities. Solid lines denote $a_{\rm k}=0.94$ and dashed lines refer to $a_{\rm k}=0.5$. For the $\Phi_{13}$ sequences in panel (b), the fixed angular momentum flux is $\mathcal{L}=2.075$ for $a_{\rm k}=0.94$ and $\mathcal{L}=2.65$ for $a_{\rm k}=0.5$, reflecting the spin dependence on the allowed angular-momentum choice for sustaining shock solutions. Note that beyond the critical limits for $\{ \mathcal{L},\Phi_{13} \}$, shock jump conditions can no longer be satisfied, and standing shocks cease to exist.}
\label{fig:shock_position}
\end{figure}

The post-shock flow is primarily characterized by the shock location and compression ratio, which we define as, $R=\rho_+H_+/{\rho_-H_-}$, where $H$ is the local disk half-thickness and the subscripts $+/-$ denote post- and pre-shock quantities, respectively. These quantities depend strongly on the conserved flow parameters $\{\mathcal{L},\Phi\}$ and the BH spin $a_{\rm k}$, and weakly on $\{\mathcal{E},F\}$ \citep{mitra2024}. In Fig.~\ref{fig:shock_position}, we therefore explore how $R$ and $r_{\rm sh}$ change when one conserved quantity is varied while the others are held fixed. We repeat this analysis for a high BH spin $a_{\rm k}=0.94$ as well as a low spin of $a_{\rm k}=0.5$. For the sequences considered, shock locations in the high-spin solutions span approximately $r_{\rm sh}\simeq4$--$110$ with $R\lesssim3.5$ (see solid curves), whereas the lower-spin solutions span approximately $r_{\rm sh}\simeq12$--$110$ with $R\lesssim2.6$ (see the dashed curve). Importantly, the resulting shock radii overlap with the reference radii for which the sound-wave crossing timescale $\tau_{c_s}(r_{\rm ref})$ and the infall/advection timescale $\tau_v(r_{\rm ref})$ overlap the observed Sgr~A* flare-duration range. For the high-spin case, these timescales span nearly the full observed duration range, while the lower-spin solutions represent a narrower range because the shocks form preferentially at larger radii. Note that we adopt a higher angular momentum flux, $\mathcal{L}=2.65$, to obtain standing-shock solutions for $a_{\rm k}=0.5$, compared with $\mathcal{L}=2.075$ for $a_{\rm k}=0.94$. This reflects the spin--orbit coupling embedded in the Kerr spacetime, for which increasing $a_{\rm k}$ shifts the allowed shock-permitting angular-momentum range toward lower $\mathcal{L}$ \citep[see] [for details]{mitra2024}.

\subsection{Energy budget estimate}
To assess whether the centrifugally supported shocked flow contains sufficient energy to account for the observed flare energetics, we construct a simple, phenomenological estimate of its available kinetic-energy reservoir. Rather than attempting to model the detailed dissipation and radiation processes, our objective is to estimate the maximum kinetic energy that could, in principle, be extracted from the shocked flow. We therefore assume that the dominant contribution originates from the reduction in the ram-pressure energy density across the shock front locally,
\begin{equation}
\delta e_{\rm sh}\propto \rho_-v_-^2-\rho_+v_+^2.
\end{equation}
Using the compression ratio, $R\simeq{\rho_+H_+}/{\rho_-H_-}$ and $[\rho u^r]=0$, the available kinetic-energy density can be approximated as
\begin{equation}
\delta e_{\rm sh}\propto \rho_-v_-^2 \left(1-\frac{\Delta H}{R} \frac{\gamma_{v_-}^2}{\gamma_{v_+}^2}\right),\qquad\Delta H\equiv\frac{H_+}{H_-},
\end{equation}
where $\gamma_v(=1/\sqrt{1-v^2})$ is the Lorentz factor. Assuming the quasi-spherical geometry of the PSC, its characteristic volume scales as $4\pi r_{\rm sh}^2H_+$. Following the unit convention, we estimate the available  kinetic-energy reservoir as
\begin{equation}
\label{eq:fluence}
E_{\rm sh}\equiv 4\pi \rho_-v_-^2 \left(1-\frac{\Delta H}{R} \frac{\gamma_{v_-}^2}{\gamma_{v_+}^2} \right) r_{\rm sh}^{2}H_+ \times M_{\rm BH}c^2.
\end{equation}

Equation~(\ref{eq:fluence}) should therefore be interpreted as an order-of-magnitude estimator of the kinetic energy available within the downstream flow, rather than as a complete model of the flare dissipation process. In particular, we neglect explicit thermal and magnetic energy reservoirs, as well as the microphysics of particle acceleration and radiative cooling. Nevertheless, magnetic fields enter the estimate through their influence on the global trans-magnetosonic solution, particularly the shock location $r_{\rm sh}$, compression ratio $R$, and the vertical extent $H_+$ of the PSC.

\section{Comparison with \textit{Chandra} data}

We compare the shock-energy estimate with the recent 25-year \textit{Chandra} flare catalog of \citet{Chandra2026}, which provides flare duration and the corresponding fluence with $2$--$10$ keV unabsorbed luminosities. To connect the stationary shock solutions to the time-domain data, we associate the observed duration with the characteristic local sound-wave crossing timescale from the shock location to the horizon. This working assumption is motivated by a picture in which part of the upstream bulk kinetic energy is available across the shock and transferred to the post-shock flow. Thus $\tau_{c_s}(r_{\rm sh})$ maps each observed duration to a shock model branch (see Fig.~\ref{fig:shock_position}) and its respective available energy $E_{\rm sh}(\tau)$. Other timescales, such as cooling, orbital, or reconnection times, may also contribute, but the sound-wave crossing time provides the primary link between the steady solutions and the observed flare durations. We retain the infall time $\tau_v$ as a complementary test in the Appendix. Neither timescale is regarded as an exact prediction of the observed flare duration. Following the catalog-quality flag provided by \citet{Chandra2026}, we separate the detected events into unflagged and flagged subsets. Flagged detections are retained in the comparison but are visually identified because their reported durations or fluences may be affected by instrumental or detection-algorithm systematics. This split provides a direct robustness check of whether the energy-budget comparison is driven by less secure long-duration events.

\begin{figure*}[t]
\centering
\includegraphics[width=0.96\textwidth]{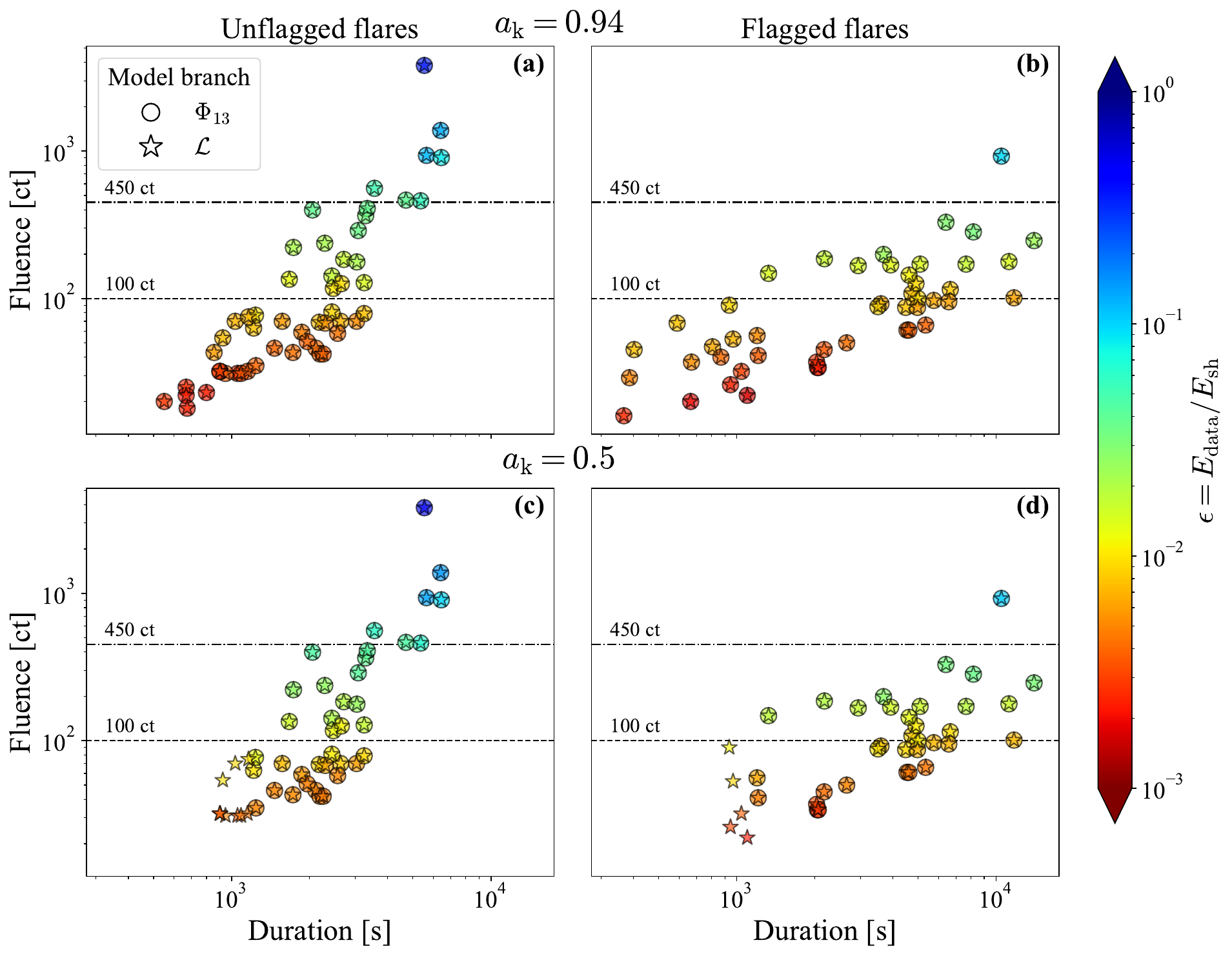}
\caption{Observed Sgr~A* X-ray flares from \citet{Chandra2026} in the duration--fluence plane, color-coded by the required efficiency $\epsilon=E_{\rm data}/E_{\rm sh}$. The duration is mapped using the sound-wave crossing timescale, $\tau_{\rm obs}\leftrightarrow\tau_{c_s}(r_{\rm sh})$. The upper panels show $a_{\rm k}=0.94$ and the lower panels show $a_{\rm k}=0.5$. The left panels (a) and (c) contain events without catalog-quality flags, while the right panels (b) and (d) contain flagged events that may be subject to instrumental and/or detection-algorithm effects, causing properties to be slightly misreported from the truth. A point is shown only when the observed duration lies within the $\tau_{c_s}$ range covered by the corresponding shock branch; the resulting numbers are listed in Table~\ref{tab:efficiency}. In each panel, circles and stars denote the $\Phi_{13}$ and $\mathcal{L}$ model branches, respectively. The common colorbar is placed outside the panels to emphasize that both catalog-quality subsets and both model branches use the same efficiency scale. Dashed and dot-dashed lines mark the weak ($<100\,{\rm ct}$), moderate ($100$--$450\,{\rm ct}$), and strong ($>450\,{\rm ct}$) flare boundaries.}
\label{fig:efficiency_duration_fluence}
\end{figure*}

The catalog count fluence is used directly in Fig.~\ref{fig:efficiency_duration_fluence} and to classify the events as weak, moderate, or strong. Because the semi-analytic model does not predict the flare spectrum nor does it fold the emission through the \textit{Chandra} instrumental response, it does not provide an
independent prediction of the observed count fluence. We instead use the measured luminosity and duration to estimate the radiated energy of each flare,
\begin{equation}
E_{\rm data}=f_{\rm bol}L_{\rm X}\tau_{\rm obs},
\end{equation}
where $L_{\rm X}$ is the observed mean $2$--$10$ keV flare luminosity reported in the catalog, $\tau_{\rm obs}$ is the flare duration, and $f_{\rm bol}$ is the bolometric correction. We adopt $f_{\rm bol}=5$, within the range $f_{\rm bol}\sim3$--$20$ inferred from broken-power-law descriptions of the Sgr~A* NIR/X-ray SED \citep{2009ApJ...698..676D,2017MNRAS.468.2447P}. This choice is also consistent with one-zone flare models constrained by GRAVITY, which place the broadband peak luminosity at $\sim1$--$5\times10^{35}\,{\rm erg\,s^{-1}}$, approximately $5$--$10$ times the $2$--$8$ keV luminosity \citep{2021A&A...654A..22G}. The precise value depends on the SED shape, the location of any synchrotron cooling break, and the possible contribution of associated submillimeter, millimeter, or radio emission. All inferred efficiencies therefore scale linearly with $f_{\rm bol}$.

For each observed duration covered by the model sequence, interpolation
along the corresponding shock branch gives
$E_{\rm sh}[r_{\rm sh}(\tau_{\rm obs})]$. We then define the required
shock-to-radiation conversion fraction as
\begin{equation}
\epsilon\equiv\frac{E_{\rm data}}{E_{\rm sh}}.
\end{equation}
Values $\epsilon>1$ are energetically incompatible with the adopted shock energy reservoir estimator, whereas $\epsilon<1$ indicates that the shocked flow contains sufficient energy in principle to power the flare. Moreover, Fig.~\ref{fig:efficiency_duration_fluence} shows that weak flares require $\epsilon\sim10^{-3}$--$10^{-2}$, moderate flares require a few percent, and the strongest events occupy the high-efficiency tail. 

\begin{table}[t]
\centering
\caption{Required conversion fractions obtained using the sound-wave crossing-time mapping $\tau_{\rm obs}\leftrightarrow\tau_{c_s}(r_{\rm sh})$. The median and maximum efficiencies are listed separately for the weak, moderate, and strong flare classes defined by \citet{Chandra2026} for the $\Phi_{13}$ and $\mathcal{L}$ solution branches at $a_{\rm k}=0.94$ and $0.5$. Here, $N$ denotes the number of detected flares in each class that fall within the duration range covered by the corresponding model branch. The values correspond to the full valid detected-flare sample; the flagged/unflagged data-quality split is shown explicitly in Fig.~\ref{fig:efficiency_duration_fluence}.}
\label{tab:efficiency}
\begin{tabular}{l|ccccc}
\hline
$a_{\rm k}$ & Branch & Class & $N$ & Median $\epsilon$ & Max. $\epsilon$\\
\hline

\multirow{6}{*}{$0.94$}
& $\Phi_{13}$ & Weak & 62
& $4.1\times10^{-3}$
& $9.7\times10^{-3}$\\

& & Moderate & 30
& $1.5\times10^{-2}$
& $3.7\times10^{-2}$\\

& & Strong & 8
& $8.7\times10^{-2}$
& $3.6\times10^{-1}$\\
\cline{2-6}

& $\mathcal{L}$ & Weak & 62
& $4.4\times10^{-3}$
& $1.0\times10^{-2}$\\

& & Moderate & 30
& $1.6\times10^{-2}$
& $3.9\times10^{-2}$\\

& & Strong & 8
& $8.5\times10^{-2}$
& $3.6\times10^{-1}$\\
\hline
\hline

\multirow{6}{*}{$0.5$}
& $\Phi_{13}$ & Weak & 34
& $5.5\times10^{-3}$
& $1.3\times10^{-2}$\\

& & Moderate & 30
& $1.8\times10^{-2}$
& $4.6\times10^{-2}$\\

& & Strong & 8
& $9.9\times10^{-2}$
& $4.1\times10^{-1}$\\
\cline{2-6}

& $\mathcal{L}$ & Weak & 48
& $5.2\times10^{-3}$
& $1.2\times10^{-2}$\\

& & Moderate & 30
& $1.9\times10^{-2}$
& $4.6\times10^{-2}$\\

& & Strong & 8
& $1.0\times10^{-1}$
& $4.2\times10^{-1}$\\
\hline

\end{tabular}
\end{table}

Table~\ref{tab:efficiency} summarizes these distributions for all valid detected flares for the high-spin $a_{\rm k}=0.94$ and lower-spin $a_{\rm k}=0.5$ cases. The tabulated values are not predicted radiative efficiencies, but the efficiencies required under Eq.~(\ref{eq:fluence}) for the fiducial bolometric correction. The two solution branches agree within a factor of $\sim2$ in both median and maximum efficiency for each flare class, demonstrating that the energy-budget conclusion is robust to the choice of which conserved quantity is varied. The unflagged subset is the cleanest observational robustness sample. Its shorter duration range removes the least secure long-duration tail but remains fully within the dynamical range relevant for the shock sequence. Therefore, the flagged/unflagged split should be viewed as a data-quality check: the robust conclusion is based primarily on the unflagged weak-to-moderate population, while flagged events are retained in Fig.~\ref{fig:efficiency_duration_fluence} for completeness.

For the unflagged subset alone, the median efficiencies for weak and moderate flares are fully consistent with the full-sample values in Table~\ref{tab:efficiency} for both $a_{\rm k}=0.94$ and $0.5$. The flagged/unflagged separation does not introduce a new flare class; it only identifies the reliability of the measured catalog properties. Since the unflagged sample spans $0.55$--$6.44\,{\rm ks}$, the high-confidence events still occupy the same dynamical window, in which the model shock branches provide available shock energies. The longest flagged events are therefore useful for completeness but are not required for the main weak-to-moderate energy-budget conclusion. With the sound-wave crossing mapping, both $a_{\rm k}=0.94$ branches cover all 100 detected flares \citep{Chandra2026}, while in the weakly spinning case ($a_{\rm k}=0.5$), $\Phi_{13}$ and $\mathcal{L}$ branches cover 72 and 86 events, respectively (Fig.~\ref{fig:efficiency_duration_fluence} and Table~\ref{tab:efficiency}). For the infall/advection mapping, the corresponding coverage becomes 99 and 97 events for $a_{\rm k}=0.94$, and 61 and 71 events for $a_{\rm k}=0.5$ (see Fig.~\ref{fig:4} and Table~\ref{tab:2} for details). For flare classes present in both mappings, the median and maximum conversion fractions differ by no more than about 15\% across the sequences considered. Thus, the choice of characteristic timescale mainly changes the range of flare durations covered by the model, while the weak-to-moderate flare energy-budget result remains largely unchanged. We therefore use $\tau_{c_s}$ for the primary comparison because it spans a broader part of the observed duration distribution, and retain $\tau_v$ as a complementary test. The more limited duration range of the $a_{\rm k}=0.5$ sequences is not interpreted as an independent constraint on the spin of Sgr~A*, since the allowed shock solutions also depend on the conserved flow parameters.

The condition $\epsilon<1$ only tests energetic allowance; physical plausibility requires that a reasonable fraction of the shock reservoir is converted into X-rays. Dissipative shocked-accretion models find maximum shock energy losses of a few percent \citep{Das-etal2022RadioShock}, so we use $\epsilon\sim0.04$ as a reference benchmark, not a universal limit. Under this stricter criterion, the model naturally accounts for weak flares and a substantial fraction of moderate flares, whereas the brightest events may require unusually efficient shock dissipation or an additional channel, such as magnetic reconnection or flux eruption. Thus, the observed duration--fluence distribution provides a useful visualization of the required conversion fractions, while the principal result is the energy-budget test of the standing shocks.

\section{Discussion and limitations}

The present work should be interpreted as an energy-budget test rather than a complete time-dependent emission model. The stationary and axisymmetric solutions used here provide characteristic shock properties, but real accretion flows around Sgr~A* are turbulent, magnetized, and time-dependent. Consequently, the inferred efficiencies should be regarded as order-of-magnitude requirements, not unique predictions for individual flare light curves.  

A central assumption is the association between the observed flare duration and the infall or sound-wave crossing time from the shock location. This choice is physically motivated because material energized during the shock transition can advect inward on a dynamical timescale comparable to observed X-ray flare durations. The sound-wave crossing time provides an alternative characteristic timescale for the same post-shock region. However, the observed duration may also be affected by cooling, orbital motion, reconnection, turbulent correlation times, or instrumental detection thresholds. The mapping $\tau_{\rm obs}\leftrightarrow\tau_v(r_{\rm sh})$ or $\tau_{\rm obs}\leftrightarrow\tau_{c_s}(r_{\rm sh})$ is therefore one of the main assumptions of the present comparison. Neither timescale is interpreted as an exact observer-frame flare duration. For flare classes represented in both mappings, the inferred conversion fractions differ by at most about 15\%, indicating that the main energy-budget conclusion is not strongly affected by the choice between these two characteristic timescales.

The inferred $E_{\rm data}$ depends on the bolometric correction $f_{\rm bol}$, which converts the observed $2$--$10$ keV luminosity into a total radiated flare energy. Because $f_{\rm bol}$ is uncertain, all quoted efficiencies scale linearly with this choice: adopting $f_{\rm bol}=2$ would reduce the efficiencies by a factor of 2.5, whereas $f_{\rm bol}=10$ would double them. Similarly, Eq.~(\ref{eq:fluence}) neglects explicit thermal and magnetic energy reservoirs. Including these contributions could modify the available energy and the effective dissipation efficiency. The density normalization is likewise tied to the fiducial accretion rate: for a fixed dimensionless flow structure, $E_{\rm sh}\propto\dot{M}$ and hence $\epsilon\propto\dot{M}^{-1}$, so the factor of a few uncertainty in the quiescent accretion rate of Sgr~A* \citep{2010ApJ...716..504S,Ressler-etal2020} propagates linearly into the required efficiencies.

The strongest flares should be interpreted with particular caution. The fact that $\epsilon<1$ indicates energetic allowance, but not necessarily that shocks alone provide the complete emission mechanism. Weak flares require small efficiencies and are naturally compatible with the shock-powered scenario. Moderate flares remain plausible, while the highest-fluence events require larger efficiencies and may involve additional magnetic processes operating on the longer timescale of $\gtrsim 5\times 10^3\,{\rm s}\sim 1.4$ hours. This interpretation is consistent with the broader observational picture in which X-ray flares are commonly associated with infrared activity, while not all infrared flares produce detectable X-ray counterparts. In the shock scenario this asymmetry arises naturally: any disturbance energetic enough to power a detectable X-ray flare also heats and compresses the inner flow, producing an infrared counterpart, whereas weaker magnetic events in the innermost region can generate infrared variability without a significant change in the shock energetics. The potentially missed ultra-weak flare population inferred by \citet{2015ApJ...799..199N} is not included as individually characterized events in the detected-flare catalog. Energetically, such events would require lower $E_{\rm data}$ and hence smaller $\epsilon$ than the detected weak flares, provided their durations remain within the dynamical range covered by the shock sequence.

Finally, the present work does not model the flare SED or its time evolution. Shock waves can in principle accelerate a nonthermal electron population, providing a plausible source of synchrotron emission in the NIR and X-ray bands. However, reproducing the observed spectral evolution, including the rising NIR spectrum, the steeper X-ray spectrum, and possible cooling-break behavior, requires a time-dependent treatment of the electron distribution, magnetic-field evolution, acceleration cutoff, and radiative cooling. We therefore regard spectral modeling as a necessary next step, while the present work is restricted to testing whether the available energy budget in shock-induced flow is sufficient to account for the observed flare energetics.

\section{Conclusions}

We have investigated whether standing shocks in low-angular-momentum magnetized accretion flows can transfer a sufficient amount of upstream bulk kinetic energy into the post-shock flow to power the Sgr~A* X-ray flares. Our main conclusions are as follows.

\begin{enumerate}
\item Global trans-magnetosonic solutions yield shock locations whose infall/advection timescale $\tau_v$ and sound-wave crossing timescale $\tau_{c_s}$ overlap the flare durations measured in the \textit{Chandra} catalog. For $a_{\rm k}=0.94$, the sound-wave crossing prescription includes all 100 detected flares for both model branches, whereas the infall/advection prescription includes 99 and 97 events for the $\Phi_{13}$ and $\mathcal{L}$ branches, respectively. The $a_{\rm k}=0.5$ solutions reproduce a narrower portion of the observed duration distribution. For flare classes common to both timescale prescriptions, the required conversion fractions differ by no more than about 15\%, indicating that the weak-to-moderate flare energy-budget result is largely insensitive to the adopted characteristic clock. Neither timescale is regarded as an exact observer-frame flare duration, and the spin comparison is not used as an independent constraint on the spin of Sgr~A* because shock formation also depends on the conserved flow parameters.

\item Within the adopted kinetic-energy estimator and fiducial bolometric correction and mass-accretion rate, the energy stored inside the shocked flow is sufficient to satisfy $\epsilon<1$ for all observed flare populations considered here.

\item Separating the catalog into unflagged and flagged detections provides a data-quality robustness check. The unflagged sample removes the least secure long-duration tail while preserving the relevant flare-duration range, so the duration agreement is not dependent on flagged events.

\item Weak flares require efficiencies of order $10^{-3}$--$10^{-2}$, while moderate flares require efficiencies of a few percent, depending on the model branch. If the radiated fraction of shock energy is limited to a few percent, the model most naturally explains weak flares and a substantial subset of moderate flares. The strongest events remain likely probes of additional magnetic dissipation physics operating on the longer timescale of $\gtrsim 5\times 10^3\,{\rm s}\sim 1.4$ hours.
\end{enumerate}

Finally, we conclude that standing shocks in low-angular-momentum accretion flows provide a physically viable bulk-kinetic energy reservoir for weak and moderate Sgr~A* X-ray flares. This result does not by itself constitute a complete flare-emission model, but it establishes that the shock-powered scenario passes a basic energetics test against the 25-year \textit{Chandra} flare catalog.

\begin{acknowledgements}
The authors thank the anonymous reviewer for the constructive feedback for improving the manuscript. SM thanks Henry Best,  Ji\v{r}\'{\i} Hor\'ak, Sebastiano von Fellenberg, Daryl Haggard, Maasaki Takahashi, Maciek Wielgus, and Bart Ripperda for their valuable comments during the preparation of this draft. In particular, we thank Daryl Haggard for a careful assessment of the comparison with the \textit{Chandra} flare catalog and the manuscript. SM also thanks MZ, BC, Tomas Ondro, JH, and W{\l}odzimierz Klu\'zniak for their hospitality during the development of this project. SM thanks especially to all the local organizing committee, the `GC crew', members of the IAU Symposium 405 at Brno. SM also thanks Pallavi Bhat and \texttt{astroplasma} group of ICTS-TIFR. This project has received funding from the European Research Council (ERC) under the European Union’s Horizon 2020 research and innovation program (grant agreement No. 951549). The Czech-Polish Mobility program of the two Academies of Sciences, titled ``Appearance and dynamics of accretion onto black holes'', is greatly appreciated. BC acknowledges the OPUS-LAP/GA\v{C}R-LA bilateral project funded by NCN (2021/43/I/ST9/01352/OPUS 22 and GF23-04053L). MZ is grateful for the support of the GA\v{C}R Junior Star grant ``Stars in galactic nuclei: interrelation with massive black holes'' no. GM24-10599M. SM thanks ``Sujata and Tatineni Prem Kumar Travel Grant Fund'' from ICTS-TIFR and `IAU Travel Grant' for supporting his travel to the IAU symposium 405 where the work was initiated and partly completed. SM acknowledges the support of the Department of Atomic Energy, Government of India, under project no. RTI4019. ZS acknowledges support from the Canadian Space Agency (23JWGO2A01 and 25JWGO4A01), the Natural Sciences and Engineering Research Council of Canada (NSERC) Discovery Grant program, the Canada Research Chairs (CRC) program, the Centre de recherche en astrophysique du Quebec, the Trottier Space Institute at McGill, and the Chalk-Rowles fellowship. 
\end{acknowledgements}

\appendix

\section{Analysis with infall timescale}

\begin{figure}[h!]
\centering
\includegraphics[width=\columnwidth]{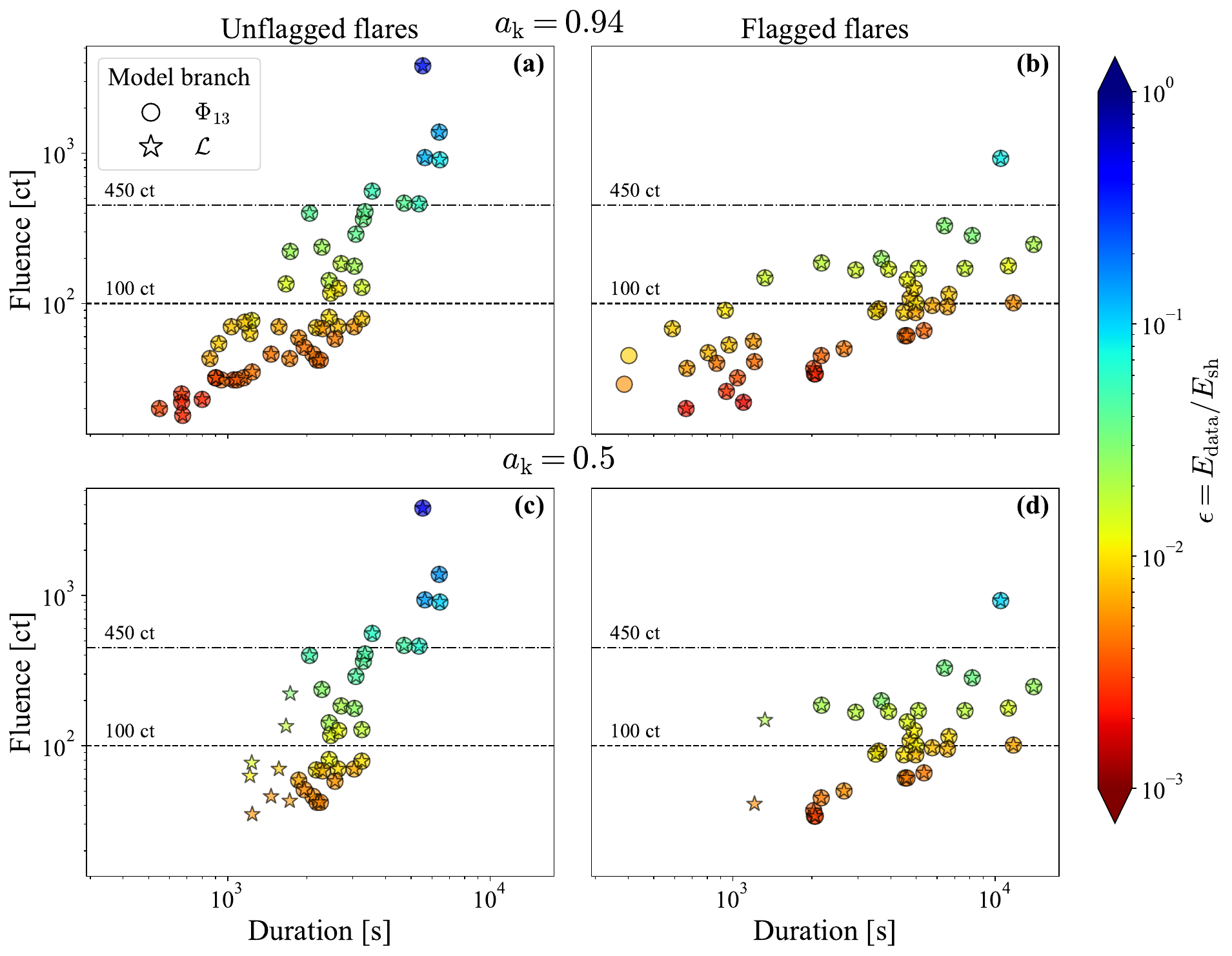}
\caption{The same catalog data and model sequences as in Fig.~\ref{fig:efficiency_duration_fluence}, but using the infall/advection-time mapping $\tau_{\rm obs}\leftrightarrow\tau_v(r_{\rm sh})$. For $a_{\rm k}=0.94$, the $\Phi_{13}$ and $\mathcal{L}$ branches cover 99 and 97 flares, respectively; for $a_{\rm k}=0.5$, they cover 61 and 71 flares. The corresponding conversion fractions are summarized in Table~\ref{tab:2}.}
\label{fig:4}
\end{figure}

Here we present the results based on the association of the observed flare duration with the infall/advection timescale. The results for the unflagged and flagged flares, two SMBH spin values ($a_{\rm k}=0.94$ and $a_{\rm k}=0.5$), and the two solution branches ($\Phi_{13}$ and $\mathcal{L}$) are shown in Fig.~\ref{fig:4}. The corresponding median and maximum shock-to-radiation conversion factors for different flare classes are summarized in Table~\ref{tab:2}, including the number of flares whose duration is matched by the corresponding solution branches.

\begin{table}[h!]
\centering
\caption{The same quantities as in Table~\ref{tab:efficiency}, but using the infall/advection-time mapping $\tau_{\rm obs}\leftrightarrow\tau_v(r_{\rm sh})$ shown in Fig.~\ref{fig:4}.}
\label{tab:2}
\begin{tabular}{l|ccccc}
\hline
$a_{\rm k}$ & Branch & Class & $N$ & Median $\epsilon$ & Max. $\epsilon$\\
\hline

\multirow{6}{*}{$0.94$}
& $\Phi_{13}$ & Weak & 61
& $4.59\times10^{-3}$
& $1.10\times10^{-2}$\\

& & Moderate & 30
& $1.57\times10^{-2}$
& $4.12\times10^{-2}$\\

& & Strong & 8
& $8.11\times10^{-2}$
& $3.56\times10^{-1}$\\
\cline{2-6}

& $\mathcal{L}$ & Weak & 59
& $4.58\times10^{-3}$
& $1.12\times10^{-2}$\\

& & Moderate & 30
& $1.63\times10^{-2}$
& $4.20\times10^{-2}$\\

& & Strong & 8
& $8.45\times10^{-2}$
& $3.76\times10^{-1}$\\
\hline
\hline

\multirow{6}{*}{$0.5$}
& $\Phi_{13}$ & Weak & 26
& $5.92\times10^{-3}$
& $1.16\times10^{-2}$\\

& & Moderate & 27
& $1.83\times10^{-2}$
& $5.0\times10^{-2}$\\

& & Strong & 8
& $9.47\times10^{-2}$
& $4.21\times10^{-1}$\\
\cline{2-6}

& $\mathcal{L}$ & Weak & 33
& $5.80\times10^{-3}$
& $1.37\times10^{-2}$\\

& & Moderate & 30
& $1.92\times10^{-2}$
& $5.02\times10^{-2}$\\

& & Strong & 8
& $9.65\times10^{-2}$
& $4.23\times10^{-1}$\\
\hline

\end{tabular}
\end{table}

\bibliography{references}{}
\bibliographystyle{aasjournalv7}

\end{document}